\documentclass[prl, twocolumn, preprintnumbers, amsmath, showpacs, amssymb, superscriptaddress, nobalancelastpage]{revtex4-2}
\usepackage{epsfig}
\usepackage{color}
\usepackage{graphicx}
\usepackage{dcolumn}
\usepackage{bm}
\usepackage{tabularx}
\usepackage[pdftex, colorlinks, citecolor=blue]{hyperref}   
\usepackage{appendix}
\usepackage{diagbox}
\usepackage{threeparttable}
\usepackage{multirow}
\usepackage{tabularx}
\usepackage{booktabs}
\usepackage{txfonts}
\usepackage{xcolor}
\usepackage{amssymb}
\usepackage{amsmath}
\usepackage{latexsym}
\usepackage{amsbsy}
\usepackage{array}
\usepackage{esvect} 
\usepackage{extarrows}
\usepackage{float}

\begin{document}

\title{Quadratic piezoelectricity from stacking-engineered interference in multilayer sliding ferroelectrics}%

\author{Jiaxuan Fan}
\affiliation{Research Center for Quantum Physics and Technologies, School of Physical Science and Technology, Inner Mongolia University, Hohhot 010021, China}

\author{Yuexing Zhao}
\affiliation{Research Center for Quantum Physics and Technologies, School of Physical Science and Technology, Inner Mongolia University, Hohhot 010021, China}

\author{Xiao-Ping Li}
\affiliation{Research Center for Quantum Physics and Technologies, School of Physical Science and Technology, Inner Mongolia University, Hohhot 010021, China}
\affiliation{Key Laboratory of Semiconductor Photovoltaic Technology and Energy Materials at Universities of Inner Mongolia Autonomous Region, Inner Mongolia University, Hohhot 010021, China}

\author{Yurong Yang}
\affiliation{National Key Laboratory of Solid State Microstructures, Nanjing University, Nanjing 210093, China}
\affiliation{Jiangsu Key Laboratory of Artificial Functional Materials, Department of Materials Science and Engineering, Nanjing University, Nanjing 210093, China}

\author{Xing-Qiu Chen}
\email{xingqiu.chen@imr.ac.cn}
\affiliation{Shenyang National Laboratory for Materials Science, Institute of Metal Research, Chinese Academy of Sciences, Shenyang 110016, China}
\affiliation{School of Materials Science and Engineering, University of Science and Technology of China, Shenyang 110023, China}

\author{Lei Wang}
\email{lwang@imu.edu.cn}
\affiliation{Research Center for Quantum Physics and Technologies, School of Physical Science and Technology, Inner Mongolia University, Hohhot 010021, China}
\affiliation{Inner Mongolia Key Laboratory of Microscale Physics and Atomic Manufacturing, Inner Mongolia University, Hohhot 010021, China}

\begin{abstract}
Designing nonlinear piezoelectricity requires suppressing the linear piezoelectric coefficient without extinguishing higher-order electromechanical response, yet a general and reconfigurable route remains lacking. Here we introduce stacking-engineered piezoelectric interference as such a mechanism in multilayer sliding ferroelectrics. Combining first-principles calculations with a generalized Ginzburg--Landau framework, we show that each interlayer gap acts as a local piezoelectric channel whose sign and magnitude are determined by stacking. Constructive interference between same-signed channels produces a linear-dominated response, whereas destructive interference between oppositely signed channels suppresses the linear coefficient while preserving a finite quadratic response. Representative MoS$_2$ and NiTe$_2$ multilayers approach the parabolic limit, with BAAC-stacked MoS$_2$ reducing the linear-to-quadratic crossover strain by a factor of 25 relative to CBA-stacked MoS$_2$. Experimentally accessible tetralayer MoS$_2$ sliding pathways further connect linear-dominated, quadratic-dominated and sign-inverted states. Here, we identify stacking-engineered interference as a design principle for programmable nonlinear electromechanics in layered materials.
\end{abstract}

\maketitle

\section{Introduction}

Piezoelectricity enables direct conversion between electrical and mechanical energy and underlies sensors, actuators, resonators and energy-harvesting devices~\cite{AEM-review-piezo,Piezoelectric_Energy_Harvesting_2022,SEZER2021105567,peh-APR-2018,SABARIANAND2020106634}.
In most materials under small strain, the polarization response is dominated by the linear piezoelectric coefficient, which determines both the magnitude and sign of the primary electromechanical coupling~\cite{FE0book-2001,Dragan_Damjanovic_1998}.
The search for high-performance piezoelectrics has therefore largely focused on maximizing this linear figure of merit~\cite{bernardini_spontaneous_1997,dutta_piezoelectricity_2021,wang_large_2023,PRB-Pe-SbBi}.
Reconfigurable electromechanics, however, requires a broader level of control: one would like to select whether the response of a single material platform is linear, nonlinear, or sign-inverted.

A particularly demanding form of such control is nonlinear piezoelectricity, where the linear term is weakened or cancelled so that higher-order electromechanical terms dominate the polarization--strain relation, producing quadratic or nearly parabolic responses~\cite{bester_importance_2006,prodhomme_nonlinear_2013,caro_origin_2015,grigoriev_nonlinear_2008}.
Recent work on epitaxially strained HfO$_2$ demonstrates that such behavior can emerge when the linear coefficient is nearly annihilated while the quadratic coefficient remains finite~\cite{cheng_tunable_2024}.
However, this compensation is achieved through material-specific competition among local structural distortions within a single bonding network, and is often tied to delicate strain, coordination, or phase-instability conditions.
A general mechanism for suppressing the linear response without relying on such critical tuning remains lacking.

The microscopic structure of the piezoelectric coefficient suggests how this problem might be reframed.
The macroscopic linear coefficient is not an elementary quantity, but an algebraic sum of contributions, commonly separated into clamped-ion and internal-strain parts~\cite{PRL_III-V,PRB-FP-CG}.
Studies of negative longitudinal piezoelectricity have shown that changing the balance among such contributions can even reverse the sign of the response~\cite{katsouras_negative_2016,liu_origin_2017,jing_understanding_2024}.
In conventional bulk crystals, however, these competing contributions are embedded in the same covalent network and are largely fixed by bonding chemistry, Born effective charges and strain-induced internal relaxations.
This makes the cancellation of the linear term difficult to design reversibly.

Multilayer van der Waals sliding ferroelectrics offer a distinct opportunity: they convert this internal competition into a spatially resolved structural degree of freedom.
Their out-of-plane polarization is governed by stacking-dependent interfacial charge redistribution~\cite{li2017binary,yang2018origin,yasuda2021stacking,stern_interfacial_2021,weston_interfacial_2022}, while vertical strain is accommodated by several mechanically compliant interlayer gaps~\cite{kim_negative_2019,you_origin_2019,qi_widespread_2021}.
Each gap can therefore act as a local piezoelectric channel whose response depends on the local stacking environment.
Because lateral sliding changes the stacking sequence without breaking covalent bonds~\cite{yasuda2021stacking,stern_interfacial_2021}, these channels can in principle be rearranged reversibly~\cite{wu_sliding_2022,Ouyang2025Electrically}.
This raises the central question of whether stacking order can make interfacial piezoelectric channels add constructively or cancel destructively, and whether sliding can switch the resulting macroscopic response between linear, nonlinear and sign-inverted regimes.

\begin{figure*}
\centering
\includegraphics[width=0.89\textwidth]{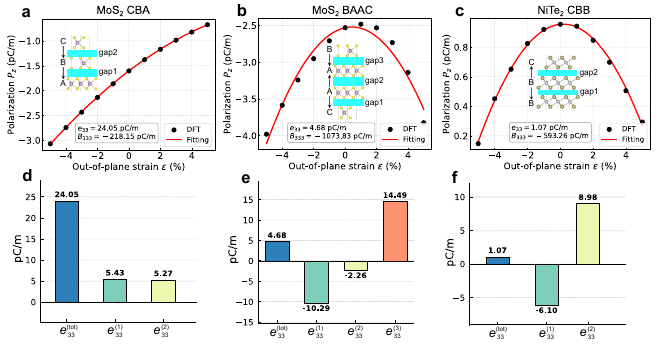}
\caption
{\textbf{Stacking-controlled constructive and destructive piezoelectric interference.}
\textbf{a--c} Calculated out-of-plane polarization \(P_z\) under vertical strain \(\epsilon\) for CBA-MoS\(_2\), BAAC-MoS\(_2\), and CBB-NiTe\(_2\). Black symbols denote first-principles data, and red curves are quadratic fits. 
Black arrows indicate the directions of local interfacial dipoles (see Supplementary Note 4).
\textbf{d--f} Total linear coefficient \(e_{33}^{(\mathrm{tot})}\) and interface-resolved response channels \(e_{33}^{(i)}\) for individual vdW gaps (see Supplementary Note 3). 
Same-signed channels in CBA-MoS\(_2\) reinforce one another, producing a large linear response, whereas opposite-signed channels in BAAC-MoS\(_2\) and CBB-NiTe\(_2\) compensate one another, suppressing \(e_{33}^{(\mathrm{tot})}\) and exposing a quadratic-dominated, near-parabolic polarization--strain response.}\label{fig1}
\end{figure*}

Here, using first-principles calculations, energy--polarization landscape analysis and a generalized Ginzburg--Landau framework, we identify stacking-engineered piezoelectric interference in multilayer van der Waals sliding ferroelectrics.
Rather than tuning a single bonding network toward a structural instability, the mechanism suggests spatial compensation can suppress the linear response among stacking-dependent interfacial channels.
Same-signed channels add constructively to yield a linear-dominated response, whereas oppositely signed channels compensate, leaving a finite quadratic response and moving the response toward the parabolic limit.
Energy and polarization landscapes rationalize the geometric origin of this cancellation: the strain-relaxation pathway becomes approximately orthogonal to the polarization gradient in interlayer-distance space.
Guided by this principle, calculated BAAC-stacked tetralayer MoS$_2$ exhibits a linear-to-quadratic crossover strain 25 times smaller than that of CBA-stacked trilayer MoS$_2$, and ferrielectric-like CBB-NiTe$_2$ provides a second near-parabolic example.
We predict that sliding pathways in tetralayer MoS$_2$ connect stacking states with linear-dominated, quadratic-dominated and sign-inverted responses.
These results support stacking-controlled piezoelectric interference as a theory-guided, testable design principle for programmable nonlinear electromechanics in van der Waals materials.

\section{Results}
\subsection{Stacking-controlled piezoelectric interference}
We first introduce a measure for the crossover from linear to quadratic piezoelectricity. Under vertical strain, the out-of-plane polarization can be expanded to second order as
\begin{equation}
    P_z(\epsilon)=P_0+e_{33}\epsilon+\frac{1}{2}B_{333}\epsilon^2 ,
\end{equation}
where \(e_{33}\) and \(B_{333}\) are the linear and quadratic piezoelectric coefficients, respectively. 
When \(e_{33}\) is suppressed but \(B_{333}\) remains finite, the response can approach a quadratic-dominated, near-parabolic form.
We characterize this regime by the vertex strain,
\begin{equation}
    \epsilon_v=\left|e_{33}/B_{333}\right|,
\end{equation}
with smaller \(\epsilon_v\) corresponding to a more nearly parabolic polarization--strain curve. 
Below we show that this condition can be achieved in multilayer sliding ferroelectrics through cancellation among stacking-dependent interfacial piezoelectric responses.

\begin{figure*}
\centering
\includegraphics[width=0.99\textwidth]{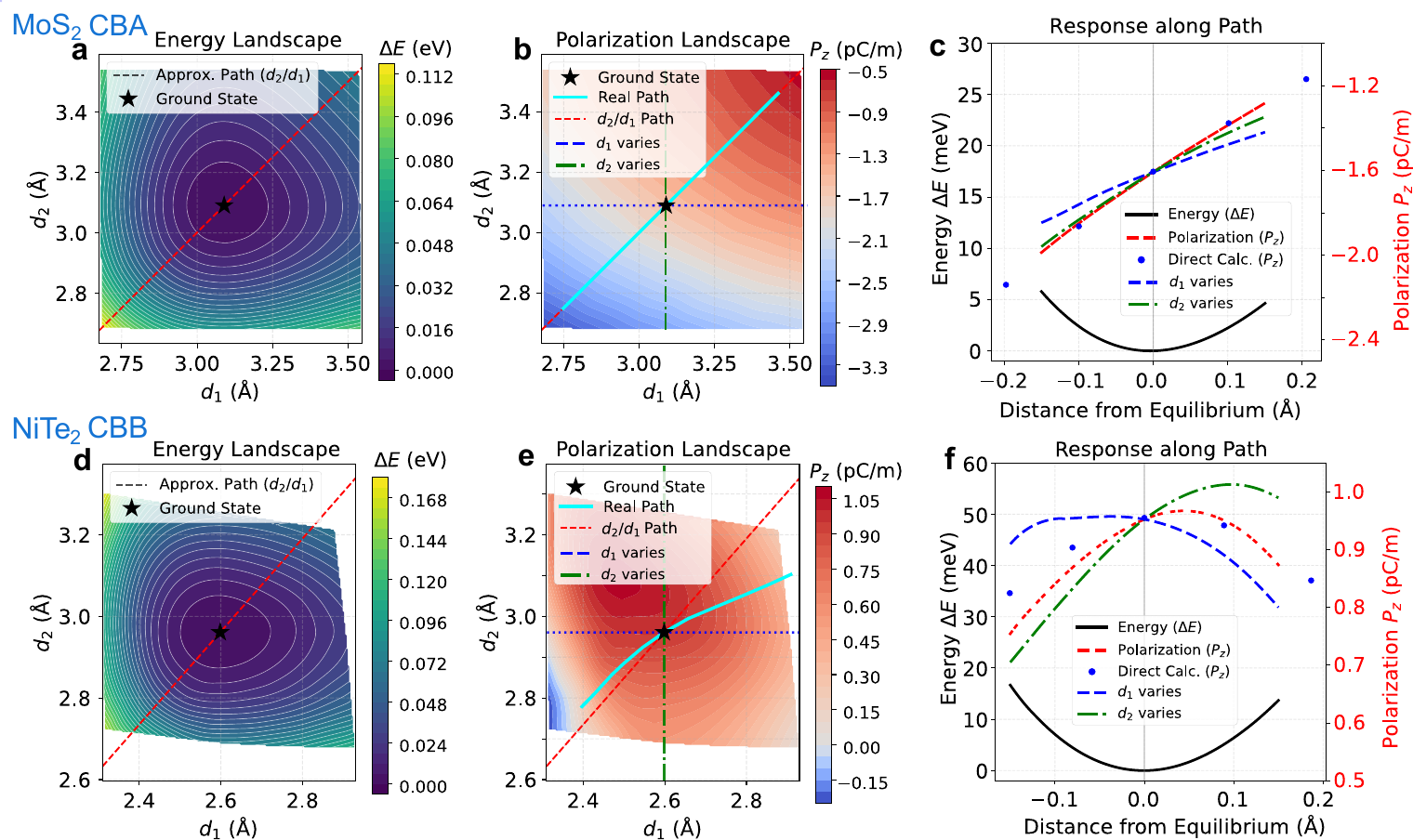}
\caption{
\textbf{Geometric origin of piezoelectric interference in energy and polarization landscapes.}
Energy and polarization landscapes are mapped in the interlayer-distance space \((d_1,d_2)\). 
\textbf{a,b} Landscapes for the constructive-interference system, CBA-stacked MoS\(_2\). 
The ground state, approximate uniform-strain path, isolated \(d_1\) and \(d_2\)  variation paths, and fully relaxed DFT strain path are indicated by a black star, red dashed line, blue and green dashed lines, and solid cyan line, respectively. 
The relaxed path crosses the polarization contours near the ground state, yielding a large polarization gradient along the strain-relaxation direction. 
\textbf{c} Energy and polarization profiles along the paths in \textbf{b}. 
The isolated \(d_1\) and \(d_2\) variation paths have same-signed slopes, revealing constructive interference between interface-resolved response channels.
\textbf{d,e} Corresponding landscapes for the destructive-interference system, CBB-stacked NiTe\(_2\). 
The relaxed path runs nearly parallel to the local polarization contours, suppressing the linear polarization change. 
\textbf{f} Energy and polarization profiles along the paths in \textbf{e}. 
The isolated \(d_1\) and \(d_2\) variation paths have opposite-signed slopes, revealing destructive interference and a nearly parabolic macroscopic response.
}\label{fig2}
\end{figure*}

Figure~\ref{fig1} presents a comparative first-principles analysis of three representative multilayer systems. 
We first consider CBA-stacked trilayer MoS$_2$ (CBA-MoS$_2$), whose out-of-plane polarization varies almost linearly with vertical strain (Fig.~\ref{fig1}a), giving a large positive linear coefficient of \(e_{33}^{(\mathrm{tot})}\approx +24.05\) pC/m and a comparatively small quadratic coefficient of \(B_{333}\approx -218.15\) pC/m. 
By contrast, BAAC-stacked tetralayer MoS$_2$ (BAAC-MoS$_2$) (Fig.~\ref{fig1}b) shows a strongly suppressed linear coefficient, \( e_{33}^{(\mathrm{tot})}\approx +4.68\) pC/m, while retaining a pronounced quadratic coefficient, \(B_{333}\approx -1073.83\) pC/m. 
A similar case is found in CBB-stacked trilayer NiTe$_2$ (CBB-NiTe$_2$) (Fig.~\ref{fig1}c), which exhibits an almost parabolic polarization--strain response with \(e_{33}^{(\mathrm{tot})}\approx +1.07\) pC/m and \(B_{333}\approx -593.26\) pC/m. 
The corresponding vertex strains decrease from \(\epsilon_v\approx 0.110\) for CBA-MoS$_2$ to \(\epsilon_v\approx 0.0044\) for BAAC-MoS$_2$ and \(\epsilon_v\approx 0.0018\) for CBB-NiTe$_2$, demonstrating a systematic crossover from predominantly linear to nearly parabolic piezoelectricity. 
Because different stacking sequences define different local environments at the vdW gaps, this contrast suggests that the macroscopic response is governed by whether the corresponding interface piezoelectric channels reinforce or compensate one another.

To diagnose the sign competition underlying these different interfacial piezoelectric responses, we perform an interface-resolved channel analysis. 
Here, \(e_{33}^{(i)}\) denotes the separately calculated out-of-plane piezoelectric response associated with interface \(i\) or the corresponding vdW gap, as defined in Supplementary Note 3. 
For CBA-MoS$_2$ (Fig.~\ref{fig1}d), the two interfacial responses have the same sign, \(e_{33}^{(1)},e_{33}^{(2)}>0\), indicating a constructive piezoelectric-interference regime in which local piezoelectric pathways mutually reinforce one another and produce a large macroscopic linear response.
For CBB-NiTe$_2$ (Fig.~\ref{fig1}f), the two interfacial responses have opposite signs, \(e_{33}^{(1)}<0\) and \(e_{33}^{(2)}>0\), providing a local signature of destructive piezoelectric interference. This compensation is associated with its ferrielectric-like stacking geometry (see Supplementary Note 4), leading to a near cancellation of the total linear coefficient.
BAAC-MoS$_2$ (Fig.~\ref{fig1}e) realizes an engineered multi-interface version of the same effect, where positive and negative interfacial contributions compensate each other and expose the otherwise hidden quadratic response.

These examples establish the central phenomenology of piezoelectric interference: stacking-dependent interfacial piezoelectric pathways can either add constructively to yield a conventional linear response or cancel destructively to reveal nonlinear piezoelectricity.

\subsection{Geometric origin of piezoelectric interference}
To clarify the microscopic origin of piezoelectric interference, we map the energy and polarization landscapes in the interlayer-distance space, \(\mathbf d=(d_1,d_2)\), where \(d_i\) denotes the spacing of the \(i\)-th vdW gap, as described in Supplementary Note 5.
The energy landscape determines how the vdW gaps relax under vertical strain, whereas the polarization landscape determines how this structural relaxation is converted into an out-of-plane polarization response. 
This representation allows a direct comparison between the constructive-interference CBA-MoS\(_2\) and the destructive-interference CBB-NiTe\(_2\).

For CBA-MoS\(_2\), the equilibrium structure lies in a nearly symmetric energy valley with \(d_1\approx d_2\) (Figs.~\ref{fig2}a,b). 
Upon vertical strain, the relaxed structure follows a coupled trajectory in the \((d_1,d_2)\) plane, shown by the cyan line. 
The internal-strain contribution to the linear piezoelectric response is governed by the directional derivative of the polarization along this relaxation path,
$dP_z/d\epsilon = \nabla_{\mathbf d}P_z \cdot d{\mathbf d}/d\epsilon$,
where \(\nabla_{\mathbf d}P_z\) is the polarization gradient in the interlayer-distance space. 
In CBA-MoS\(_2\), the relaxation path cuts across the polarization contours and is nearly aligned with \(\nabla_{\mathbf d}P_z\) near equilibrium. 
The large projection gives a robust linear polarization response (Fig.~\ref{fig2}c). 
Moreover, the polarization change along the fully relaxed path is larger than those obtained by varying \(d_1\) or \(d_2\) separately, indicating that the two interfacial responses reinforce each other under strain.

CBB-NiTe\(_2\) shows the opposite geometric relation. 
Its ferrielectric-like state has unequal interlayer distances, \(d_1\neq d_2\), and the strain-relaxation path runs nearly along the local polarization contours (Figs.~\ref{fig2}d,e). 
Equivalently, the relaxation direction is almost orthogonal to the polarization gradient near equilibrium, $\left.  \nabla_{\mathbf d}P_z\cdot   \frac{d\mathbf d}{d\epsilon}     \right|_{\epsilon=0}    \simeq 0 $. 
This near-orthogonality suppresses the linear term in \(P_z(\epsilon)\), resulting in a nearly parabolic polarization--strain relation, consistent with the first-principles result in Fig.~\ref{fig2}f.

The landscape further reveals how the cancellation arises. 
The polarization profile extracted along the relaxed coupled path reproduces the direct first-principles polarization--strain curve, showing that the interlayer-distance landscape captures the essential electromechanical response. 
Near equilibrium, the relaxed trajectory lies between two limiting paths in which only \(d_1\) or only \(d_2\) is varied. 
These limiting paths have opposite linear polarization slopes, corresponding to interfacial piezoelectric contributions of opposite sign. 
The actual relaxation therefore samples competing local piezoelectric pathways, leading to near algebraic cancellation of the macroscopic linear coefficient.

Thus, constructive and destructive piezoelectric interference can be understood geometrically as the alignment or near-orthogonality between the strain-relaxation direction and the polarization gradient in the interlayer-distance landscape. 
This microscopic picture explains the first-principles phenomenology in Fig.~\ref{fig1} and motivates a more general description in terms of stacking-dependent coupling parameters.

\begin{figure*}
\centering
\includegraphics[width=0.99\textwidth]{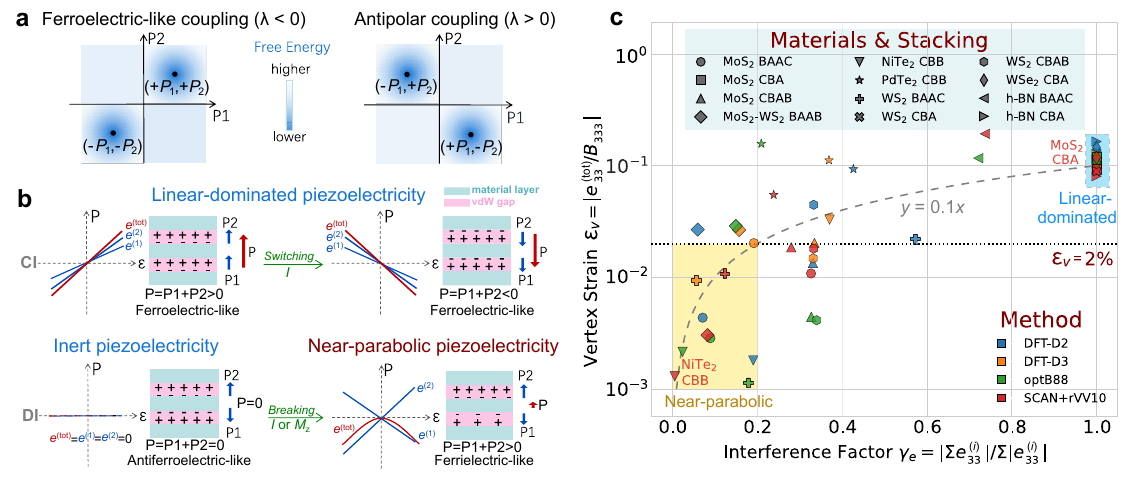}
\caption{
\textbf{Ginzburg--Landau description and design space for piezoelectric interference.}
\textbf{a} Schematic Ginzburg--Landau free-energy landscapes for two coupled out-of-plane polar subunits. 
Ferroelectric-like coupling favors parallel polar alignment, whereas antipolar coupling favors antiparallel alignment. 
\textbf{b} Schematic classification of constructive-interference (CI) and destructive-interference (DI) responses.
Ferroelectric-like stacks yield linear-dominated CI responses whose sign reverses upon polarization switching. 
Antiferroelectric-like stacks can exhibit symmetry-enforced cancellation and become piezoelectrically inert. 
Ferrielectric-like stacks retain a finite net polarization while enabling opposite-signed local response channels to compensate, suppressing \(e_{33}^{(\mathrm{tot})}\) and revealing a near-parabolic DI response.  
\textbf{c} First-principles materials map using the interference factor \(\gamma_e\) and vertex strain \(\epsilon_v\). 
Large-\(\gamma_e\) systems are linear-dominated, whereas small-\(\gamma_e\), small-\(\epsilon_v\) systems fall in the near-parabolic regime. 
Colors denote vdW treatments, and symbols denote material and stacking configurations. 
The stacking configurations and piezoelectric data are provided in Supplementary Notes 7 and 8.
}\label{fig3}
\end{figure*}

\subsection{Ginzburg--Landau framework and materials design space}
To move beyond the material-specific landscape picture, we formulate a minimal Ginzburg--Landau description for coupled interfacial piezoelectric units (see Supplementary Note 6 for details). 
The model is intended to capture the local electromechanical response around a non-centrosymmetric equilibrium stacking configuration.
As illustrated in Fig.~\ref{fig3}a, a multilayer vdW structure can be simplified into two effective out-of-plane polar subunits, \(P_1\) and \(P_2\), associated with distinct interfacial environments. 
The corresponding free-energy density is written as
\begin{equation}
\begin{aligned}
    G &= \sum_{i=1,2}
    \left(
    \frac{A_i}{2}P_i^2+\frac{\beta_i}{4}P_i^4
    \right)
    -\sum_{i=1,2} g_i \epsilon P_i  \\
    &\quad +\lambda P_1P_2
    +\frac{C}{2}\epsilon^2
    +\mathcal{O}(\epsilon P_i^2,\ldots),
\end{aligned}
\label{eq:landau_full}
\end{equation}
where \(A_i\) and \(\beta_i\) describe the local polar stiffness, \(g_i\) is the strain-induced piezoelectric driving field, and \(\lambda\) denotes the effective coupling between the two polar subunits. 

We expand the polarizations as
\begin{equation}
    P_i(\epsilon)=P_{i,0}+e^{(i)}\epsilon+\mathcal{O}(\epsilon^2),
\label{eq:Pi_expansion}
\end{equation}
where \(e^{(i)}=dP_i/d\epsilon|_{\epsilon=0}\). The projected model-level linear piezoelectric coefficient is then given by
\begin{equation}
    e^{(\mathrm{tot})}
    =
    e^{(1)}+e^{(2)}.
\end{equation}
This algebraic balance between the two GL linear-response channels leads to the interference-based classification summarized in Fig.~\ref{fig3}b, in which the local piezoelectric responses either reinforce or compensate one another, corresponding to constructive and destructive interference, respectively.

To determine \(e^{(1)}\) and \(e^{(2)}\), we linearize the equilibrium conditions \(\partial G/\partial P_i=0\) around the equilibrium state, giving
\begin{equation}
    \begin{pmatrix}
        \alpha_1 & \lambda \\
        \lambda & \alpha_2
    \end{pmatrix}
    \begin{pmatrix}
        e^{(1)} \\ e^{(2)}
    \end{pmatrix}
    =
    \mathbf g^{\mathrm{eff}}.
\label{eq:matrix_in_main}
\end{equation}
Here, \(\alpha_i=A_i+3\beta_iP_{i,0}^2\) is the harmonic stiffness around \(P_{i,0}\), and \(\mathbf g^{\mathrm{eff}}\) includes the effective strain-driving terms after linearization. 

In the ferroelectric-like coupling regime depicted in Fig.~\ref{fig3}b, \(\lambda<0\) favors parallel alignment of the two polar subunits. 
For same-signed effective local strain-driving terms, \(g_1^{\mathrm{eff}}g_2^{\mathrm{eff}}>0\), the ferroelectric-like coupling reinforces the two response channels, so that the local linear responses share the same sign and add constructively,
\begin{equation}
    e^{(1)}e^{(2)}>0,\qquad
    |e^{(\mathrm{tot})}|=|e^{(1)}|+|e^{(2)}|.
\end{equation}
This constructive addition keeps \(e_{33}^{(\mathrm{tot})}\) large, so achieving a small vertex strain,
\(\epsilon_v=|e_{33}^{(\mathrm{tot})}/B_{333}|\), would require a large \(B_{333}\).
Although sizeable second-order coefficients can occur in both conventional semiconductors~\cite{bester_importance_2006,prodhomme_nonlinear_2013,caro_origin_2015} and van der Waals systems such as CBA-MoS$_2$, the nonlinear contribution generally remains subdominant near zero strain when a strong linear channel persists. 
This observation motivates a destructive-interference strategy that directly suppresses the total linear response through compensation between opposite-signed local response channels.

By contrast, in the antipolar coupling regime depicted in Fig.~\ref{fig3}b, \(\lambda>0\) favors antiparallel alignment of the two polar subunits.
In an approximately antipolar ferrielectric-like configuration, \(P_{1,0}P_{2,0}<0\).
When the effective strain-driving terms are opposite-signed, \(g_1^{\mathrm{eff}}g_2^{\mathrm{eff}}<0\), and their stiffness-weighted contributions nearly cancel, the local linear responses become opposite-signed and strongly compensate:
\begin{equation}
    e^{(1)}e^{(2)}<0,\qquad 
    |e^{(\mathrm{tot})}| \ll |e^{(1)}|+|e^{(2)}|.
\end{equation}
This destructive-interference regime provides an algebraic route to suppressing the linear piezoelectric response through compensation analogous to that in antiferroelectric-like configurations, while retaining a finite net polarization.

Importantly, the approximately antipolar arrangement that can support linear compensation does not impose the same cancellation on the quadratic response. Within the Ginzburg--Landau framework, the second-order response is governed by distinct source terms, including contributions proportional to \(\beta_iP_{i,0}[e^{(i)}]^2\), together with higher-order strain-coupling and local-stiffness contributions (see Supplementary Note 6). 
Quadratic cancellation would therefore require an additional condition, which may be symmetry-enforced in an ideal antiferroelectric-like limit. 
Ferrielectric-like stacks retain inequivalent interfaces and a finite net polarization, so quadratic cancellation is not generally enforced. Consistently, \(B_{333}\) remains finite in CBB-NiTe$_2$ and BAAC-MoS$_2$ even when opposite-signed interface-resolved linear responses strongly suppress \(e_{33}^{(\mathrm{tot})}\), accounting for their near-parabolic piezoelectric responses.

\begin{figure*}
\centering
\includegraphics[width=0.95\textwidth]{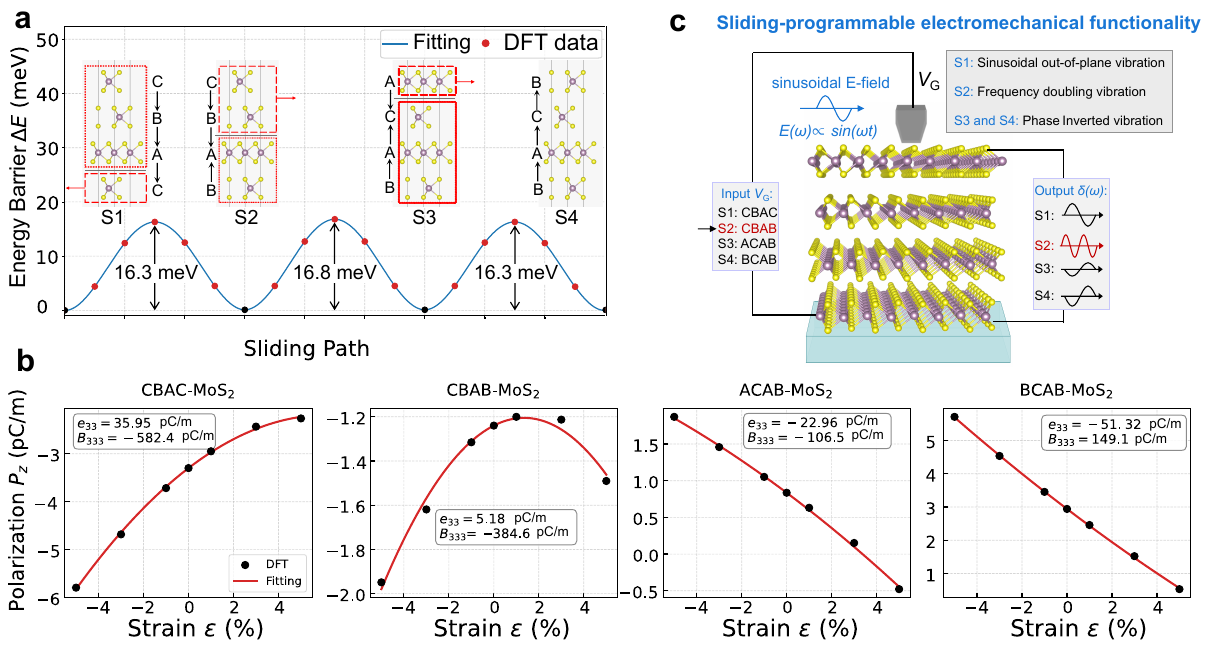}
\caption{
\textbf{Sliding-programmable piezoelectric interference in tetralayer MoS$_2$.}
\textbf{a} Minimum-energy sliding pathway connecting four metastable stacking states, CBAC, CBAB, ACAB, and BCAB, labeled S1--S4. 
Insets show the corresponding atomic structures, with black arrows indicating local interfacial dipoles.
\textbf{b} Polarization--strain responses of the four sliding states. 
S1 exhibits a constructive-interference, linear-dominated response, whereas S2 realizes a destructive-interference state with a suppressed linear coefficient and dominant quadratic response. 
S3 and S4 recover linear responses with opposite signs, reflecting reversal of the effective polar axis. 
\textbf{c} Conceptual operating principle of sliding-programmable electromechanical functionality. 
By selecting different non-volatile stacking states, the same tetralayer is predicted to access linear-dominated, quadratic-dominated, and phase-inverted response states; the quadratic-dominated state provides a potential route to frequency doubling.}
\label{fig4}
\end{figure*}

To examine whether this interference picture extends beyond the representative examples, we performed a comparative first-principles survey of 12 multilayer vdW systems. 
For each candidate, the response was evaluated using four treatments of vdW interactions, including DFT-D2, DFT-D3, optB88-vdW, and SCAN+rVV10 (see Supplementary Note 7). Although the absolute coefficients vary with the dispersion treatment, the constructive/destructive classification and low-\(\gamma_e\) trend remain consistent across the four treatments (Supplementary Note 8).
We quantify the degree of piezoelectric interference by the dimensionless factor
\begin{equation}
    \gamma_e =
    \frac{\left|\sum_i e_{33}^{(i)}\right|}
    {\sum_i \left|e_{33}^{(i)}\right|},
\label{eq:gamma}
\end{equation}
where \(e_{33}^{(i)}\) is the interface-resolved response channel defined in Supplementary Note 3.
A value of \(\gamma_e \approx 1\) corresponds to constructive addition of local responses, whereas \(\gamma_e \approx 0\) indicates destructive cancellation.

Figure~\ref{fig3}c maps the calculated systems using the interference factor \(\gamma_e\) and the vertex strain \(\epsilon_v\). 
A clear trend emerges: systems with smaller \(\gamma_e\) generally show smaller \(\epsilon_v\), indicating a polarization--strain response closer to a zero-centered parabola. 
In particular, candidates with \(\gamma_e<0.2\) exhibit vertex strains of only a few percent, demonstrating that cancellation of the linear interfacial response is an effective descriptor for parabolic piezoelectricity.

The resulting design map distinguishes linear-dominated responses from near-parabolic candidates within the surveyed systems.
CBA-MoS\(_2\)-type systems cluster in the constructive-interference regime with \(\gamma_e\approx 1\), where the polarization response is dominated by the linear term. 
By contrast, CBB-NiTe\(_2\) and MoS\(_2\) stackings such as BAAC and CBAB fall into the destructive-interference regime. 
The reported realization of ferrielectric-like CBAB MoS\(_2\) provides experimental access to a relevant sliding ferroelectric platform~\cite{wu_sliding_2022}; the electromechanical response states predicted here remain to be directly tested.
Together, the model and materials map support stacking-controlled compensation of interfacial piezoelectric pathways as a theory-guided route to suppress the macroscopic linear coefficient while retaining the quadratic response in the surveyed systems.

\subsection{Sliding-programmable piezoelectric interference}
Having identified \(\gamma_e\) as a descriptor of stacking-dependent interference, we next examine computationally whether different interference regimes can be accessed within a single material.
Multistate sliding ferroelectrics provide a natural platform for this purpose, because lateral translation changes the sequence of interfacial environments without changing the chemical composition. 
We investigate this possibility in tetralayer MoS\(_2\), where several metastable stacking configurations are connected by ferroelectric sliding.

Figure~\ref{fig4}a shows the minimum-energy sliding pathway connecting four stacking states, denoted S1--S4, corresponding to CBAC, CBAB, ACAB, and BCAB stackings. 
These states are close in energy and separated by moderate calculated sliding barriers, comparable to energy scales reported for van der Waals sliding ferroelectrics~\cite{yasuda2021stacking,stern_interfacial_2021,weston_interfacial_2022,wu_sliding_2022,Ouyang2025Electrically}. 
Because each stacking state defines a distinct set of interfacial piezoelectric pathways, sliding provides a direct way to move the same material through different interference regimes.

The calculated polarization--strain responses indeed change qualitatively along this pathway (Fig.~\ref{fig4}b). 
In the S1 state, CBAC-stacked MoS\(_2\) lies in the constructive-interference regime, with \(\gamma_e\approx 1\), and shows a nearly linear response with positive \( e_{33}^{(\mathrm{tot})} \). 
After sliding to the S2 state, corresponding to ferrielectric-like CBAB stacking, inequivalent interfacial piezoelectric pathways largely compensate one another. 
This destructive-interference state has a strongly suppressed linear coefficient and a response dominated by the quadratic term. 
Further sliding to S3 and S4 restores a predominantly linear response, but with the opposite sign of \( e_{33}^{(\mathrm{tot})} \), reflecting reversal of the effective polar axis.

This sequence suggests a conceptual route to electrically programmable electromechanical functionality, as illustrated in Fig.~\ref{fig4}c. 
In a state-select--then-probe protocol, a lateral switching field would select a non-volatile stacking state, while a small vertical ac field would probe the electromechanical response within that state. 
The constructive-interference states are predicted to provide linear electromechanical transduction, with outputs proportional to the input field. 
The destructive-interference state suppresses the linear channel, so that within a quasistatic quadratic-response approximation, the response would contain a second-harmonic mechanical component~\cite{stanton_nonlinear_2010,yamamoto_nonlinear_2000}. 
The oppositely polarized linear states are predicted to provide a phase-inverted response. 
Thus, the calculations suggest that a tetralayer MoS\(_2\) stack could be programmed among linear, quadratic-response, and phase-inverted modes by selecting its sliding configuration.

This functionality goes beyond ordinary ferroelectric reversal. 
Polarization reversal can invert the sign of piezoelectric coefficients that are odd under reversal, thereby switching between positive and negative linear responses. 
By contrast, the frequency-doubling state requires a more restrictive ferrielectric-like stacking in which inequivalent interfacial piezoelectric pathways nearly cancel, suppressing \( e_{33}^{(\mathrm{tot})} \) while retaining a finite nonlinear coefficient. 
The calculated sliding pathway of tetralayer MoS\(_2\) satisfies this condition, suggesting that stacking-engineered piezoelectric interference can provide both a static descriptor for nonlinear-piezoelectric candidates and a testable degree of freedom for reconfigurable electromechanics in multilayer sliding ferroelectrics.

In summary, first-principles calculations, polarization--strain analysis, energy--polarization landscape mapping, and a Ginzburg--Landau description identify piezoelectric interference as a mechanism for tuning the balance between linear and quadratic electromechanical responses in the multilayer van der Waals systems studied here. 
The macroscopic out-of-plane piezoelectric response can be rationalized as the superposition of stacking-dependent interfacial response pathways: constructive addition is associated with linear-dominated behavior, whereas near compensation suppresses the linear coefficient while retaining a finite quadratic response.

This interference picture provides a design logic based on the internal balance among interfacial electromechanical pathways. 
The interference factor \(\gamma_e\) provides a compact descriptor for distinguishing constructive and destructive regimes within the surveyed multilayer stackings.
In multistate sliding ferroelectrics, this balance can be further controlled dynamically: lateral sliding changes the stacking sequence and thereby switches the same material among distinct electromechanical response modes. 
More broadly, the concept motivates analogous tests in layered systems in which multiple polar or interfacial units contribute comparably to the macroscopic response, including multilayer sliding ferroelectrics, moiré superlattices~\cite{PRL-moire-FE,Zhang2024-moire,Zheng2020-moire}, hybrid layered materials~\cite{Hu2025-2Dhyb}, and oxide heterostructures~\cite{Liu2024-PZO_FRI,PRAMANIK2025116426}. 
Our results therefore suggest an interferometric strategy for designing nonlinear and reconfigurable piezoelectric materials by controlling the cooperative or compensating contributions of internal piezoelectric channels.

\section{Methods}

\subsection{First-principles calculations} 
All first-principles calculations were performed based on density functional theory (DFT) using the projector augmented-wave (PAW) method ~\cite{kresse1999ultrasoft}, as implemented in the Vienna \textit{ab initio} Simulation Package (VASP) ~\cite{kresse1996efficient,kresse1996efficiency}. The generalized gradient approximation (GGA) in the form of the Perdew-Burke-Ernzerhof (PBE) functional was adopted for the exchange-correlation potential ~\cite{perdew1996generalized}. The kinetic energy cutoff for the plane-wave basis set was consistently set to 500 eV. The Brillouin zone was sampled using a Monkhorst-Pack $\Gamma$-centered $k$-point mesh with a density of $21 \times 21 \times 1$ for the unit cells of typical transition metal dichalcogenides (TMDs), ensuring energy convergence. To eliminate spurious interactions between periodic images, a vacuum layer of at least 15 \AA\ was applied along the out-of-plane direction. All atomic structures were fully relaxed until the residual forces on each atom were less than 0.01 eV/\AA, and the convergence criterion for the electronic self-consistent field loop was set to $10^{-6}$ eV. To assess sensitivity to the dispersion description, we repeated the calculations using four vdW treatments: (1) the semiempirical DFT-D2 method ~\cite{grimme2006semiempirical}; (2) DFT-D3 with Becke-Johnson damping ~\cite{grimme2010consistent, grimme2011effect}; (3) the non-local optB88-vdW functional ~\cite{klimes2011van}; and (4) the meta-GGA SCAN+rVV10 functional ~\cite{sun2015strongly, sabatini2013nonlocal}. The absolute coefficients depend on the treatment, whereas the CI/DI classification is evaluated for consistency in Supplementary Note 8. The minimum energy pathways and energy barriers for ferroelectric sliding were determined using the Climbing Image Nudged Elastic Band (CI-NEB) method ~\cite{NEB}. 

\subsection{Polarization and piezoelectricity} 
The out-of-plane polarization ($P_z$) was calculated by integrating the total charge density along the surface normal direction. For a slab in a supercell with area $A$, $P_z = \frac{1}{A} \left( \sum_I Z_I^{\text{val}} z_I - \int z \rho(\mathbf{r}) d^3\mathbf{r} \right)$, where $Z_I^{\text{val}}$ and $z_I$ are the valence charge and coordinate of ion $I$, and $\rho(\mathbf{r})$ is the electron density. We utilized this classical dipole method because it is uniformly applicable to both semiconducting (e.g., MoS$_2$) and semimetallic (e.g., NiTe$_2$) systems, whereas the Berry phase formalism is ill-defined for the latter ~\cite{ding2017prediction, yang2018origin}; the treatment of semimetallic NiTe$_2$ is discussed in Supplementary Note 1~\cite{PhysRevB-FEM-pi-bilayers, shi_two-dimensional_2020}. We strictly benchmarked this approach against the modern Berry phase theory for semiconducting bilayers (h-BN and MoS$_2$) and observed excellent agreement across the entire strain range (see Supplementary Note 1). The piezoelectric coefficients were derived by fitting the DFT data via formula $P_z(\epsilon) = P_0 + e_{33}\epsilon + \frac{1}{2} B_{333}\epsilon^2$ (for more details see Supplementary Note 1).

\section*{Data Availability}
The data supporting the findings of this work are available within the article and Supplementary Information. All data are available from the corresponding authors on reasonable request.

\section*{Code Availability}
All DFT calculations were performed with VASP, which is proprietary software for which the Wang’s lab owns a license.

\section*{Acknowledgements}
This work was financially supported by the National Natural Science Foundation of China (Grants No. 52188101, No. 12304165, No. 12304086, and No. 12564017), the Natural Science Foundation of Inner Mongolia Autonomous Region (Grant No. 2026QB016), the ``Grassland Talents” project of the Inner Mongolia autonomous region (Grant No. 21200-242920), and the Startup Project of Inner
Mongolia University (Grant No. 21200-5223733).

\section*{Author Contributions}
L.W. designed research; J.F. performed research; L.W., J.F. and Y.Z. analyzed data; X.-Q.C. supervised the project; and J.F., Y.Z., X.-P.L., Y.Y., X.-Q.C., and L.W. wrote the paper.

\section*{Competing Interests}
The authors declare no competing interests.

\end{document}